 \documentclass[twocolumn]{aastex62}

 \hypersetup{linkcolor=red,citecolor=blue,filecolor=magenta,urlcolor=cyan}

\shorttitle{Colors of Contact Binaries}
\shortauthors{Audrey Thirouin and Scott S. Sheppard}

\begin{document}
 
\title{Colors of Trans-Neptunian Contact Binaries}

\correspondingauthor{Audrey Thirouin}
 \email{thirouin@lowell.edu}

\author[0000-0002-1506-4248]{Audrey Thirouin}
\affil{Lowell Observatory, 1400 W Mars Hill Rd, Flagstaff, Arizona, 86001, USA.}

\author[0000-0003-3145-8682]{Scott S. Sheppard}
\affiliation{Department of Terrestrial Magnetism (DTM), Carnegie Institution for Science, 5241 Broad Branch Rd. NW, Washington, District of Columbia, 20015, USA.}



\begin{abstract}

The g'r'i' colors of seven likely and potential contact binaries in the Kuiper belt were acquired with the Magellan-Baade telescope and combined with colors from the literature to understand contact binary surfaces. The likely and potential contact binaries discovered in the dynamically Cold Classical population display very-red/ultra-red colors. Such a color is common in this sub-population and infers that the Cold Classical contact binaries were formed in-situ. The likely contact binaries found in several mean motion resonances with Neptune have colors from moderately to ultra-red suggesting different formation regions. Among the nine contact binaries discovered in resonances, five have very-red/ultra-red colors and four have moderately-red surfaces. Based on the very-red/ultra-red colors and low to moderate inclination of the contact binaries in resonances, these contact binaries are maybe escaped dynamically Cold Classicals that are now trapped in resonances. Moderately-red surfaces are common in diverse sub-populations of the Kuiper belt and thus pinpointing their origin is difficult though they are most likely captured objects formed in the giant planet area. Finally, for the contact binary population we report an anti-correlation between inclination and g'-r', as noticed in the rest of this belt. We also have hints for trends between eccentricity, perihelion distance, rotational period and g'-r', but as we are still dealing with a limited sample, additional data are required to confirm them. 
\end{abstract}



\section{Contact Binaries} 
\label{sec:intro}

Contact and close binaries are found in abundance across the Solar System in most of the small body populations \citep{Agarwal2017, Ryan2017, Massironi2015, Benner2015, Mann2007}. In the Kuiper belt, the existence of tight binaries was suspected but undiscovered as they are below the resolution of the \textit{Hubble Space Telescope} \citep{Porter2012}. But, \citet{Sheppard2004} found that the lightcurve of 2001~QG$_{298}$ was compatible with a contact binary configuration. They also estimate that the percentage of contact binaries should be $\sim$30$\%$. Because of a significant change in amplitude for the lightcurve of 2001~QG$_{298}$, \citet{Lacerda2011} modeled the system and concluded that if contact binaries have a similar large obliquity, we had underestimated their fraction. 

Recently, several likely and potential contact binaries were found in the Kuiper region through lightcurve studies \citep{Lacerda2014, ThirouinSheppard2017, Thirouin2017, Thirouin2018, Thirouin2019, Rabinowitz2018, Rabinowitz2019}. The definition of likely, confirmed and potential contact binaries is available in \citet{Thirouin2019}. To summarize, a confirmed contact binaries displays a lightcurve amplitude larger than 0.9~mag and a U-/V-shape morphology whereas a likely contact binary has also a large amplitude, but below the 0.9~mag threshold. Also, in \citet{Thirouin2019}, we report several potential contact binaries showing a large lightcurve amplitude but unfortunately without a full lightcurve allowing us to classify them as likely/confirmed contact binaries, and thus to be conservative we consider them as potential contact binaries as more data are needed to classify them. Finally, the flyby of 2014~MU$_{69}$ by the \textit{NASA New Horizons} spacecraft and its stellar occultation confirmed that this dynamically Cold Classical trans-Neptunian object (TNO) is also a contact binary \citep{Stern2019, Moore2018}. Currently, the census of contact binaries is: 
\begin{itemize}
\item In the dynamically Cold Classical population: one confirmed, 2014~MU$_{69}$, two likely, 2002~CC$_{249}$ and 2004~VC$_{131}$, and two potential, 2004~MU$_{8}$ and 2004~VU$_{75}$ \citep{ThirouinSheppard2017, Thirouin2019}. 
\item In the 3:2 resonance (or Plutino population): one confirmed, 2001~QG$_{298}$, and four likely, 2014~JQ$_{80}$, 2014~JO$_{80}$, 2014~JL$_{80}$ and 2000~GN$_{121}$ \citep{Thirouin2018, Lacerda2011, Sheppard2004}. 
\item In the 2:1 resonance (or twotino): one likely, 2012~DX$_{98}$ \citep{ThirouinPrep}.
\item In the 7:4 resonance: Manw\"{e}-Thorondor is a resolved wide binary and to interpret the mutual event season, it has been argued that Manw\"{e} (i.e., the primary) is a contact binary \citep{Rabinowitz2017, Rabinowitz2018}. 2013~FR$_{28}$ is also a likely contact binary based on its large amplitude partial lightcurve \citep{ThirouinPrep}.
\item In the 5:2 resonance: one likely, 2004~TT$_{357}$ \citep{Thirouin2017}.
\item In the Haumea family: one likely, 2003~SQ$_{317}$ \citep{Lacerda2014}.
\end{itemize}
To summarize, the contact binary population can be up to fifteen TNOs: five dynamically Cold Classical, nine resonant TNOs and one in the Haumea family. Based on the number of objects observed for large amplitude lightcurves, \citet{Thirouin2018} find that 40-50$\%$ of the Plutinos are equal-sized contact binaries while \citet{Thirouin2019} found only 10-25$\%$ of Cold Classicals are likely near-equal sized contact binaries. This suggests that dynamical evolution is important in contact binary formation. 

So far, we have studied the contact binary rotational properties but in this work, we will focus on their surface colors. The Kuiper belt has a variety of colors from neutral to ultra-red objects \citep{Peixinho2015, Marsset2019}. Some sub-populations are showing some specific color characteristics. The dynamically Cold Classical population with mostly only very-red/ultra-red \citep{Peixinho2004}. The resonances 7:4 and 5:3 within the main Kuiper belt are mostly ultra-red. The 3:2 and 2:1 resonances, just interior and exterior of the Kuiper belt are a mixture of colors. The 5:2, 3:1 and more distant resonances show mostly only moderately red colors \citep{Sheppard2012}. The very-red/ultra-red colors indicate an in-situ formation beyond Neptune whereas the mixture of colors suggest that objects formed in the inner and outer Solar System are now trapped in some resonances \citep{Sheppard2012, Fraser2012}. 
 
Following, we will report new colors for seven contact binaries and summarize the already published colors of seven more. We will use colors to pinpoint where these contact binaries may have formed, and we will look for tendencies between colors and orbital elements that may differentiate the contact binaries from the other TNOs.


\section{Observations and Reduction}

Data were obtained with the Magellan-Baade telescope at Las Campanas Observatory (Table~\ref{ObsLog}). For all the observing runs, the wide-field imager named IMACS was used. IMACS has eight 2k$\times$4k CCDs and a field of view of 27.4$\arcmin$. 

When observing an object for colors, it is crucial to take into account its lightcurve \citep{Schwamb2018}. In fact, the lightcurve amplitude correction will affect the colors in case of moderate to large variability objects. To minimize such an effect, images were taken consecutively and thus the amplitude correction is not needed. Rotational periods of the objects in this work are between 6 and 35~h (some periods are still to be determined, but seem to be longer than 10~h). Our exposure times were between 300 and 500~s depending on filters and the brightness of the objects and thus we are only covering a small fraction of the full rotation during the color measurements.

Every night, several biases and dome flats (in g', r' and i' filters) were taken to create a median flat (in each filter) and a median bias to correct the science images. Standard fields from \citet{Smith2005} were also imaged for absolute calibration purposes. Finally, the growth curve technique was used to select the optimal aperture radius \citep{Howell1989}. Our procedure is detailed in \citet{Thirouin2012}.

\begin{deluxetable*}{lcccc|cc|ccc}
\tablecaption{\label{ObsLog} Observing circumstances are available in this table. All runs made use of the Sloan filters (g'r'i'). Two columns are for the object's dynamical classification and inclination. The last seven objects have colors already published (i.e., no observing log to report). The case of Manw\"{e}-Thorondor ((385446) 2003~QW$_{111}$) will be discussed in Section 4. }
\tablewidth{0pt}
\tablehead{
TNO  & UT-date  &   r$_h$ &  $\Delta$ & $\alpha$  & Dyn.  & i & g'-r' & g'-i' &Reference \\
                      &  [MM/DD/YYYY]   &  [AU]  &  [AU]  &  [$^{\circ}$]   & Class.& [$^\circ$] & [mag] & [mag] &}
\startdata
\hline
 2004~VC$_{131}$ & 12/12/2018  & 40.776  & 39.852 & 0.5  &  Cold & 0.5& 0.89$\pm$0.06 &1.40$\pm$0.05& This work\\
 2004~VU$_{75}$ & 12/12/2018  & 43.667  & 42.933& 0.9  &  Cold   & 3.3& 0.96$\pm$0.06 &1.42$\pm$0.05& This work\\
 2012~DX$_{98}$ & 03/01/2019  &  34.376 & 35.123 & 1.1& 2:1 & 13.1 & 0.85$\pm$0.06 & 1.25$\pm$0.06 &This work\\
  2013~FR$_{28}$ & 03/02/2019  &   33.577 &  34.253 &1.2  & 7:4 & 3.0& 0.95$\pm$0.05 & 1.36$\pm$0.05 & This work\\
 2014~JL$_{80}$ & 02/03/2019   &  28.471 & 28.749 &1.9 & 3:2& 6.2& 0.74$\pm$0.05 & - & This work\\
 2014~JO$_{80}$ & 05/17/2018  & 32.192  & 31.231 & 0.6& 3:2&  15.7& 0.67$\pm$0.03 & 0.91$\pm$0.03&This work\\
  2014~JQ$_{80}$ &  03/03/2019  &  31.686  &31.850   &  1.8 & 3:2&8.0 & 0.76$\pm$0.05 & 1.05$\pm$0.05 & This work\\  
  \hline
  \hline
 2004~TT$_{357}$ & -  & -  &- &  - & 5:2  & 9.0 & 0.74$\pm$0.03 &0.99$\pm$0.04& \citet{Thirouin2017}\\
(139775) 2001~QG$_{298}$$^{a}$ & -  & -  &- &  - & 3:2  &6.5 & 0.80$\pm$0.03 & - & \citet{Sheppard2004}\\
(47932) 2000~GN$_{171}$$^{a}$ & -  & -  &- &  - & 3:2  & 10.8 & 0.80$\pm$0.04 & 1.21$\pm$0.04 &\citet{Sheppard2002}\\
 (385446) 2003~QW$_{111}$ & -  & -  &- &  - & 7:4   & 2.7& 0.85$\pm$0.06 & 1.20$\pm$0.05& \citet{Sheppard2012}\\
(126719) 2002~CC$_{249}$ & -  & -  &- &  - & Cold  & 0.8 & 0.97$\pm$0.06& 1.24$\pm$0.05&\citet{ThirouinSheppard2017}\\
(486958) 2014~MU$_{69}$$^{a}$ & -  & -  &- &  - & Cold   & 2.5 & 0.95$\pm$0.14& 1.42$\pm$0.14 & \citet{Benecchi2018}\\ 
 2004~MU$_{8}$ & -  &-   &- &  - & Cold  & 3.6& 1.15$\pm$0.17  & - & \citet{Petit2011} \\
  2003~SQ$_{317}$$^{a}$ & -  & -  &- &  - & Haumea   & 28.6 & 0.46$\pm$0.18&- & \citet{Lacerda2014}\\ 
\enddata
\tablenote{Objects with colors in the BVRI filters. \citet{Smith2002} used to convert to the Sloan filter system.}
\end{deluxetable*}


\section{Color Results: New and Published}
\label{sec:colors}

As displayed in Figure~\ref{fig:Colors}, we first report new colors of seven likely and potential contact binaries discovered by \citet{Thirouin2018, Thirouin2019}. Second, we will summarize the already published colors of five likely/confirmed contact binaries. In this section, we will focus on contact binaries identified through lightcurves studies. The special cases of 2014~MU$_{69}$ and Manw\"{e}-Thorondor will be discussed in the next Section.  

\subsection{New Cold Classical Contact Binary Colors}

\paragraph{2004~VC$_{131}$} This dynamically Cold Classical rotates every 15.7~h and has a variability of 0.55~mag \citep{Thirouin2019}. Its colors are ultra-red and are typical of the Cold Classical population: g'-r'=0.89$\pm$0.06~mag and g'-i'=1.40$\pm$0.05~mag.

\subsection{New Resonant Contact Binary Colors}
 
\paragraph{2014~JL$_{80}$} With a periodicity of about 35~h, the Plutino 2014~JL$_{80}$ is the slowest likely contact binary reported so far \citep{Thirouin2018}. We imaged 2014~JL$_{80}$ and computed: g'-r'=0.74$\pm$0.05~mag. This color suggests a moderately red surface.  

\paragraph{2014~JO$_{80}$} \citet{Thirouin2018} identified 2014~JO$_{80}$ as a likely Plutino contact binary. Based on data obtained on May 17$^{th}$, 2018, we derive moderately red colors: g'-r'=0.67$\pm$0.03~mag, r'-i'=0.24$\pm$0.03~mag and g'-i'=0.91$\pm$0.03~mag. 
 
\paragraph{2014~JQ$_{80}$} \citet{Thirouin2018} reported 2014~JQ$_{80}$ as a likely contact binary in the Plutino population. Based on images obtained in March 2019, we estimate moderately red colors: g'-r'=0.76$\pm$0.05~mag, and g'-i'=1.05$\pm$0.05~mag. 

\paragraph{2012~DX$_{98}$} This 2:1 resonant object has a periodicity of 20.80~h, a variability of 0.47~mag, and its lightcurve morphology suggests a contact binary configuration \citep{ThirouinPrep}. Its colors are very-red: g'-r'=0.85$\pm$0.06~mag and g'-i'=1.25$\pm$0.06~mag. 
\\

\paragraph{2013~FR$_{28}$} This resonant 7:4 object is a likely contact binary based on its variability of about 0.8~mag in a few hours \citep{ThirouinPrep}. Based on data carried out in March 2019, we find an ultra-red color: g'-r'=0.95$\pm$0.05~mag, and g'-i'=1.36$\pm$0.05~mag. 

\subsection{New and Published Target of Interest Colors}

The next two objects, 2004~MU$_{8}$ and 2004~VU$_{75}$, were identified as targets of interest based on their large lightcurve amplitudes over a few hours \citep{Thirouin2019}. 

\paragraph{2004~MU$_{8}$} \citet{Petit2011} reported some color information for 2004~MU$_{8}$. More details are available regarding these colors at the Besan\c{c}on photometric database for Kuiper-Belt Objects and Centaurs \footnote{\url{https://bdp-obs.utinam.cnrs.fr/src_base/download.php}}. There are 4 g'-bands, 4 r'-bands and 1 i'-band. Unfortunately, the data are from several nights and the g' and r' bands are not consecutive, so lightcurve amplitude can be an issue. The median in g' is 6.82~mag and the median in r' is 5.66~mag, so g'-r'=1.15~mag suggesting that 2004~MU$_{8}$ is ultra-red. But further observations should be obtained to confirm this color.

\paragraph{2004~VU$_{75}$} This target of interest displays a variability of $\sim$0.4~mag over few hours. \citet{Thirouin2019} presented several potential lightcurves but did not favor any option. Colors of 2004~VU$_{75}$ are: g'-r'=0.96$\pm$0.06~mag and g'-i'=1.42$\pm$0.05~mag. This ultra-red color is typical of the Cold Classical population.    \\

 \subsection{Published Contact Binary Colors}
 
Several confirmed/likely contact binaries have color results already published.  

\paragraph{2002~CC$_{249}$} This dynamically Cold Classical TNO was identified as a likely contact binary by \citet{ThirouinSheppard2017}. They found an ultra-red color of g'-r'=0.97$\pm$0.06~mag, and g'-i'=1.24$\pm$0.05~mag. 
 
\paragraph{2001~QG$_{298}$} \citet{Sheppard2004} classified this Plutino as a contact binary based on its lightcurve morphology and the extreme amplitude of 1.14~mag. \citet{Lacerda2011} obtained a second lightcurve with a smaller amplitude of 0.7~mag. The amplitude change is compatible with a system with a large obliquity observed at different viewing geometries. \citet{Sheppard2004} estimated the following colors$\footnote{The conversion to the Sloan system gives: g'-r'=0.80$\pm$0.03~mag.}$: V-R=0.60$\pm$0.02~mag and B-V=1.00$\pm$0.04~mag. There is no I-band available for this object.  

\paragraph{2000~GN$_{171}$} \citet{Sheppard2002} reported the first lightcurve of the Plutino 2000~GN$_{171}$ with a rotation of 8.3~h for a variability of 0.61~mag. The second lightcurve obtained by \citet{Dotto2008} confirmed such a find. \citet{Lacerda2007} inferred that the lightcurve can be due to a triaxial ellipsoid or due to a contact binary. Recently, we favored the contact binary option (see \citet{Thirouin2018} for details). Colors$\footnote{We calculated: g'-r'=0.80$\pm$0.04~mag and g'-i'=1.21$\pm$0.04~mag \citep{Smith2002}.}$ obtained by several teams are in agreement: V-R=0.63$\pm$0.03~mag, B-V=0.92$\pm$0.04~mag, and R-I=0.56$\pm$0.03~mag \citep{Sheppard2002, Boehnhardt2002, Doressoundiram2007, Tegler2016}.

\paragraph{2004~TT$_{357}$} \citet{Sheppard2012} collected a large color dataset for objects in Neptune's resonances. 2004~TT$_{357}$ was one of the 5:2 resonant in this sample. \citet{Sheppard2012} calculated the following colors: g'-r'=0.74$\pm$0.03~mag, and g'-i'=0.99$\pm$0.04~mag.

\paragraph{2003~SQ$_{317}$} \citet{Lacerda2014} obtained a large amplitude lightcurve (0.85~mag) with a period of 7.21~h for this object. They also measured a B-R=1.05$\pm$0.18~mag. 2003~SQ$_{317}$ belongs to the Haumea family$\footnote{g'-r'=0.46$\pm$0.18~mag using the previously mentioned conversion}$ \citep{Lacerda2014, Snodgrass2010}.

  \begin{figure*}
\includegraphics[width=19cm, angle=0]{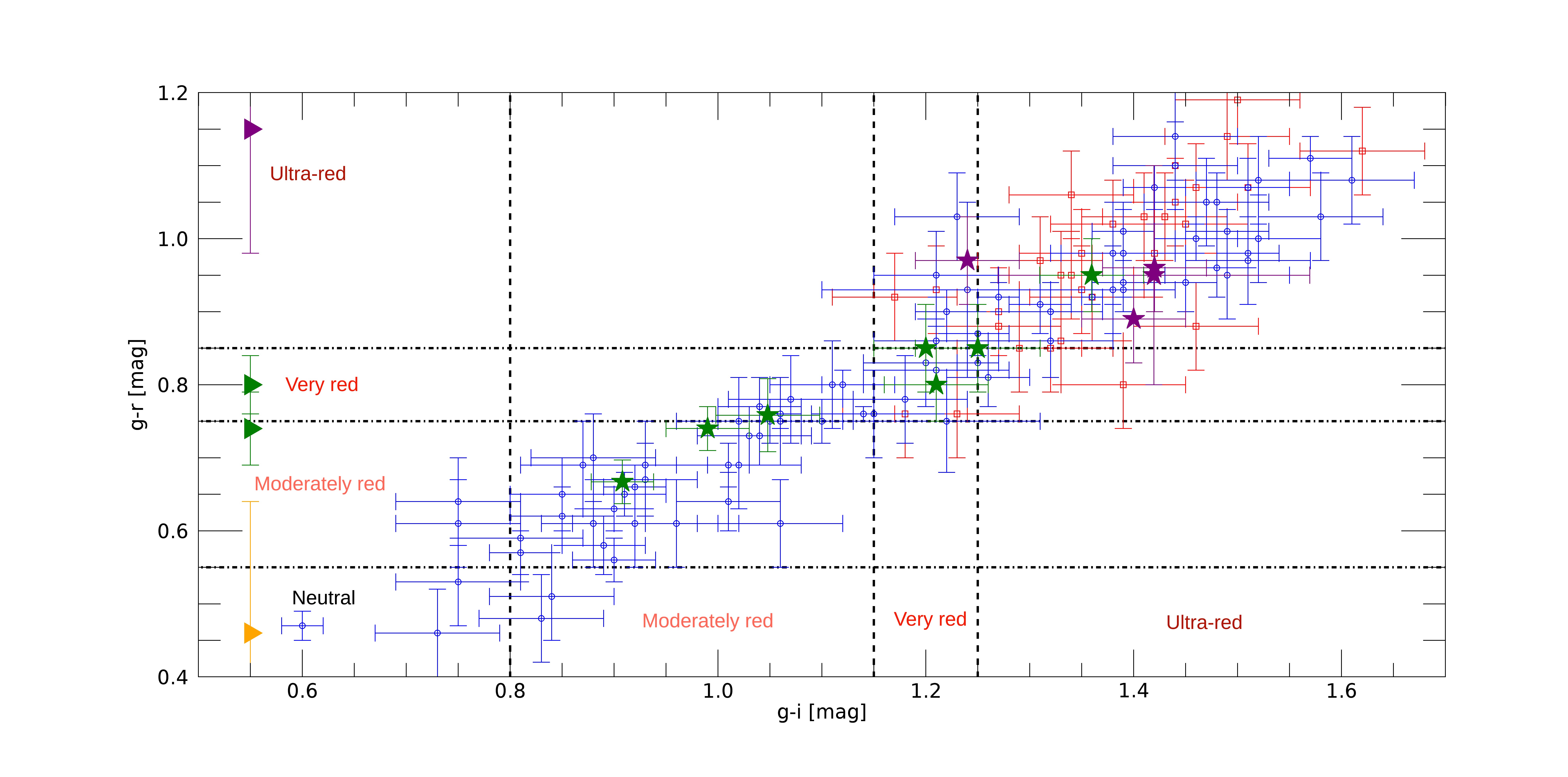}
 \caption{Colors of resonant (blue open circles) and dynamically Cold Classical TNOs (red open squares) are plotted for non-contact binary. Values are from \citet{Sheppard2012, Peixinho2015}, and references therein. Dynamically Cold Classicals discussed in this work as potential/likely contact binaries are indicated with a purple star whereas the resonant potential/likely contact binaries are plotted with a green star. As the i-bands for 2001~QG$_{298}$, 2003~SQ$_{317}$, 2004~MU$_{8}$, and 2014~JL$_{80}$ are not available, only the g'-r' is plotted (green right triangles for resonant objects, orange for the Haumea family member, and purple triangle for the Cold Classical). In the case of 2004~MU$_{8}$, there is one image in the i'-band but for reasons detailed in Section~\ref{sec:colors}, we are not considering it. Manw\"{e}-Thorondor and 2014~MU$_{69}$ are included in this plot.}
\label{fig:Colors}
\end{figure*}


\section{Discussion}
\label{sec:analysis}

  \begin{figure*}
  \includegraphics[width=20cm, angle=0]{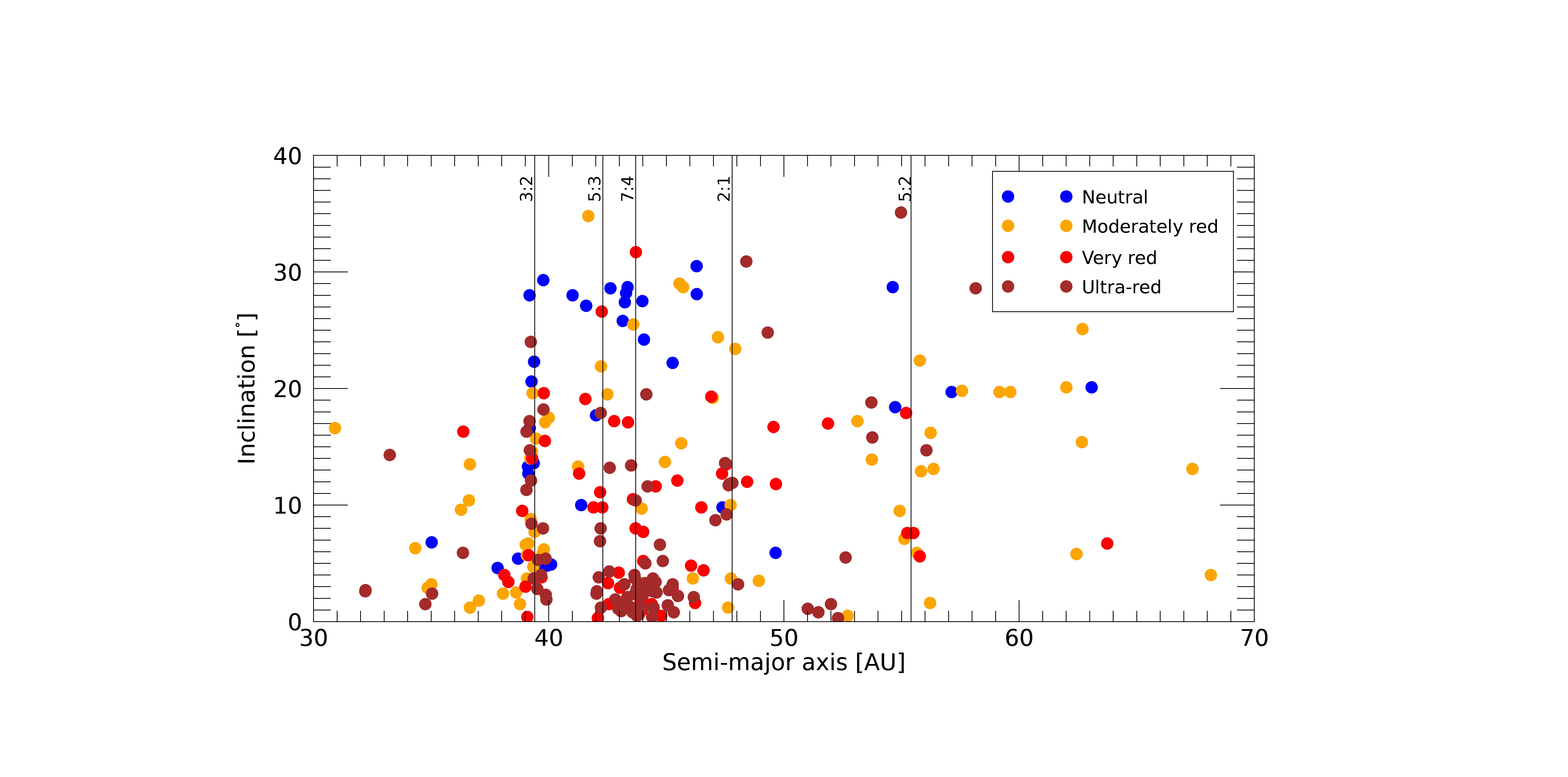}
\includegraphics[width=20cm, angle=0]{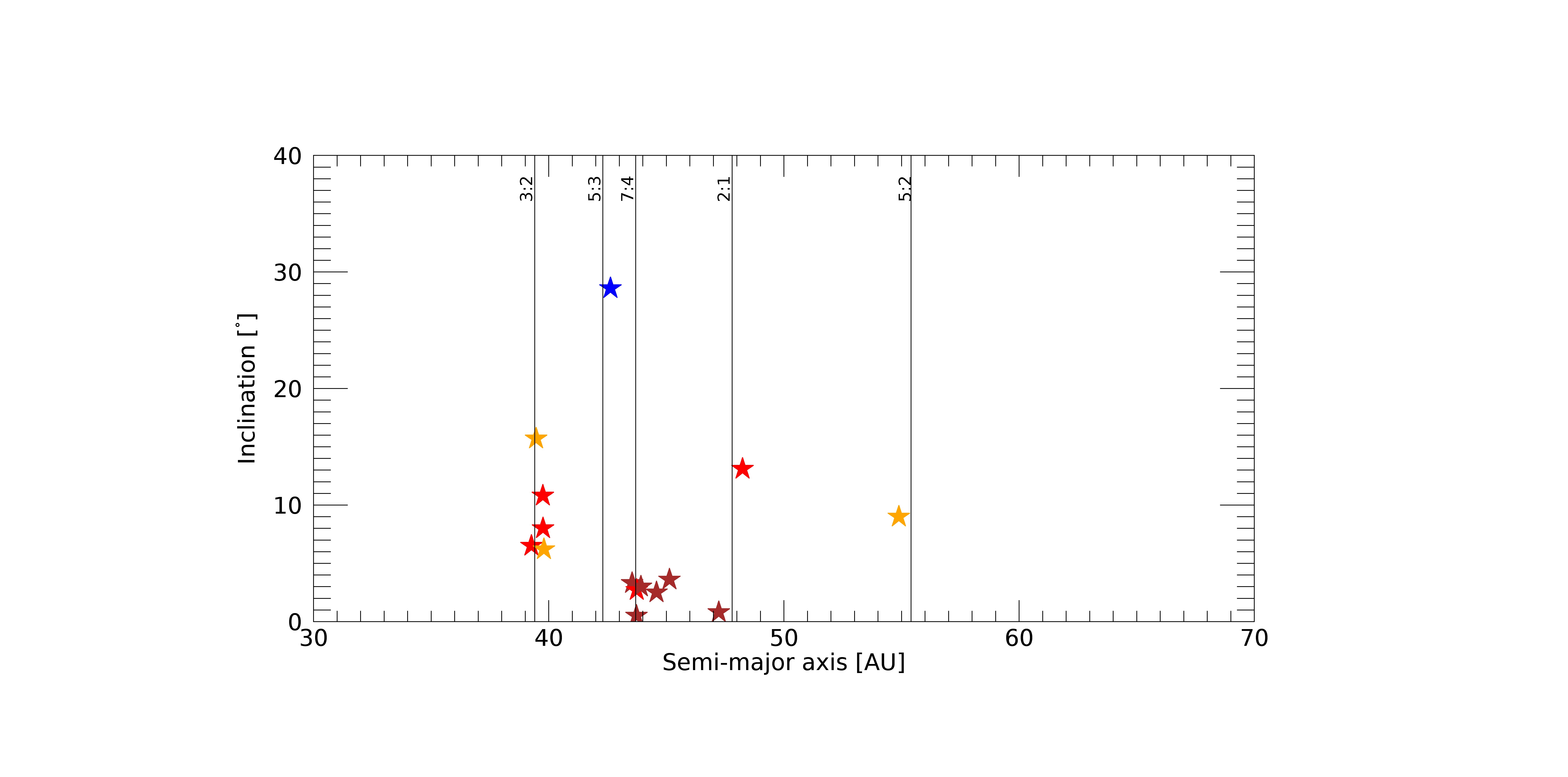}
 \caption{\textit{Color distribution in the Kuiper belt}: We plotted all TNOs with known colors (upper plot). Only objects with a semi-major axis from 30 to 70~AU are reported for clarity. Stars in the lower plot represent the contact binaries reported in this work. Same color code used for both plots.     }
\label{fig:Color2}
\end{figure*}
 
 \subsection{The Dynamically Cold Classical population} 
 
So far, two likely contact binaries (2004~VC$_{131}$ and 2002~CC$_{249}$) and two potential contact binaries (2004~MU$_{8}$ and 2004~VU$_{75}$) have been identified in the dynamically Cold Classical population through lighcurve studies \citep{Thirouin2019, ThirouinSheppard2017}. To this list, we have to add 2014~MU$_{69}$ which based on the \textit{New Horizons} flyby is a contact binary \citep{Stern2019}. Colors of 2014~MU$_{69}$ were obtained before the flyby with the \textit{HST} (\textit{Hubble Space Telescope}). \citet{Benecchi2018} reported a very red surface for this object with a F606W-F814W=1.03$\pm$0.14~mag$\footnote{\citet{Benecchi2018} converted this set of filters to V-I=1.35~mag.}$ using the \textit{HST} set of filters. Using \citet{Benecchi2018} and \citet{Smith2002}, we derived g'-i'=1.42$\pm$0.14~mag and g'-r'=0.95$\pm$0.14~mag for 2014~MU$_{69}$. This very red color was confirmed by the flyby data \citep{Stern2019}. 

In Figure~\ref{fig:Colors}, color results for these five contact binaries are plotted. They all\footnote{2014~MU$_{69}$ is close to our very-red/ultra-red separation, but we want to emphasize that we had to convert the \citet{Benecchi2018} color estimate to g'-r' and thus the uncertainty is larger than wanted.} have a g'-r'$>$0.85~mag suggesting that their surfaces are ultra-red. This color is typical of the dynamically Cold Classical TNOs, which may be further distinguished by their color through z-band photometry \citep{Pike2017}. It is thought that the dynamically Cold Classicals have been formed roughly where they are now, and they never suffered any catastrophic collisional processes \citep{Batygin2011}. With an in-situ formation, they are considered the most pristine objects among the TNOs and their surfaces are also likely primordial. As our likely/potential contact binaries and 2014~MU$_{69}$ also display the typical color of the rest of the dynamically Cold Classical TNOs, they have likely also formed in-situ, and thus are not interlopers in this population.  

 \subsection{The 3:2, 2:1, 7:4 and 5:2 resonances}
 
Neptune's mean motion resonances are interesting to understand Neptune's migration towards the outer Solar System as well as constrain if the migration was smooth or grainy \citep{Nesvorny2015}. \citet{Sheppard2012} studied in detailed the color distribution of resonant TNOs using new data and the literature. Thirteen resonances were studied in \citet{Sheppard2012}, but here we will only discuss the resonances where contact binaries have been discovered (i.e., 3:2, 2:1, 7:4 and 5:2). The dynamically Cold Classical population is ``delimited'' by the 3:2 (at 39.4~AU) and 2:1 resonances (47.8~AU). The 3:2 and 2:1 have a large variety of colors suggesting that these resonances have trapped objects from different areas of the Solar System. The distant 5:2 resonance at 55.4~AU is dominated by moderately-red objects unlike the 7:4 (at 43.7~AU) within the main classical Kuiper belt that has only ultra-red objects \citep{Sheppard2012}. 
 
So far, only one TNO in the 5:2 resonance is classified as likely contact binary: 2004~TT$_{357}$ with moderately-red color, similar to the color found for most 5:2 objects. 

In the 2:1, 2012~DX$_{98}$ is a likely contact binary with a very-red surface. In the 3:2 resonance, five likely/confirmed contact binaries have been found: 2001~QG$_{298}$ and 2000~GN$_{171}$ with very-red surfaces, and 2014~JL${80}$, 2014~JQ$_{80}$, and 2014~JO$_{80}$ with a moderately-red surface so the 3:2 contact binaries show a range of colors like the general 3:2 population. 

In the 7:4, 2013~FR$_{28}$ displays an ultra-red color. Lots of ultra-red TNOs have been found at low inclination in the 5:3 and 7:4 resonances. Such a find is not surprising as these resonances are near the dynamically Cold Classical population, at 42.3~AU and 43.7~AU. These ultra-red objects were likely formed in the Cold Classical population but are now trapped in resonances (i.e., escaped Cold Classicals). 

Based on their color survey, \citet{Sheppard2012} also found low-inclination ultra-red objects in inner and outer resonances suggesting that escaped Cold Classicals can be trapped further out and in from their main reservoir between the 3:2 and 2:1 resonances. Once trapped in resonances, objects can dynamically diffuse and end up with higher inclination. Therefore, finding escaped Cold Classical at higher inclination is possible. 

The 2:1 object, 2012~DX$_{98}$, could be an escaped Cold Classical based on its very red color.
2013~FR$_{28}$ is likely from the Cold Classical population based on its ultra-red color. The very red colors of the Plutinos 2000~GN$_{171}$ and 2001~QG$_{298}$ make them possible candidates of escaping the Cold Classical region. 

Four of our likely contact binaries, 2014~JL$_{80}$, 2014~JO$_{80}$, 2014~JQ$_{80}$, and 2004~TT$_{357}$ have moderately red surfaces and thus are not linked to the dynamically Cold Classical population. Moderate colors are typical of the Scattered Disk, the Detached Objects and the Hot Classicals, and thus we are not able to pinpoint the exact origin of these three likely contact binaries \citep{Sheppard2012}. 

Finally, we highlight the case of Manw\"{e}-Thorondor (aka (385446) 2003~QW$_{111}$). This object is a resolved wide-binary in the 7:4 resonance at low inclination \citep{Grundy2014}. In order to interpret the lightcurve of this system as well as its mutual event season, it is proposed that the primary, Manw\"{e}, is a contact binary \citep{Rabinowitz2018, Rabinowitz2019}. The colors of the system have been obtained by \citet{Sheppard2012}: g'-r'=0.85$\pm$0.06~mag, g'-i'=1.20$\pm$0.05~mag, and r'-i'=0.40$\pm$0.04~mag. These values are consistent with \citet{Grundy2014} estimates using HST data. Therefore, Manw\"{e}-Thorondor displays a very-red surface close to the ultra-red region and thus making it likely from the Cold Classical population. 

In conclusion, there are six ultra-red, five very-red, three moderately-red and one neutral contact binaries. 

\subsection{The Haumea family}

The Haumea family was discovered as a cluster of objects with similar icy surface composition and proper orbital elements by \citet{Brown2007}. Over the years, several objects have been added to the family and it was also argued that rockier members could be part of the family too \citep{Volk2012, Trujillo2011, Snodgrass2010, Schaller2008, Ragozzine2007, Brown2007}. Lightcurves have been analyzed for the members and candidates of the family \citep{Thirouin2016, Hastings2016, Lacerda2014, Carry2012}. One of the confirmed member, 2003~SQ$_{317}$ is a likely contact binary and was characterized by \citet{Lacerda2014}. This object has the typical neutral to blue color of the rest of the icy family members. It is unclear how this object formed, mainly because it it still unclear how the family was formed. Rotational fission, graze and merge impact, catastrophic collision have been proposed but none of these model is able to fully match the observables \citep{CampoBagatin2016, Ortiz2012, Leinhardt2010, Schlichting2009, Brown2007}. As the history of this family is different than the rest of the TNOs, this family should be treated as a separate case. The formation of 2003~SQ$_{317}$ as contact binary is likely different than the formation of the other contact binaries.  

\subsection{Colors versus Orbital Elements, Sizes and Rotational Periods.}
 
 First of all, we emphasize that the number of contact binaries in the Kuiper belt is still very limited, and thus all the trends presented in this sub-section have to be taken with caution. Discovering and characterizing more contact binaries are needed to improve our understandings of this population.  

 \begin{figure*}
\includegraphics[width=9cm, angle=0]{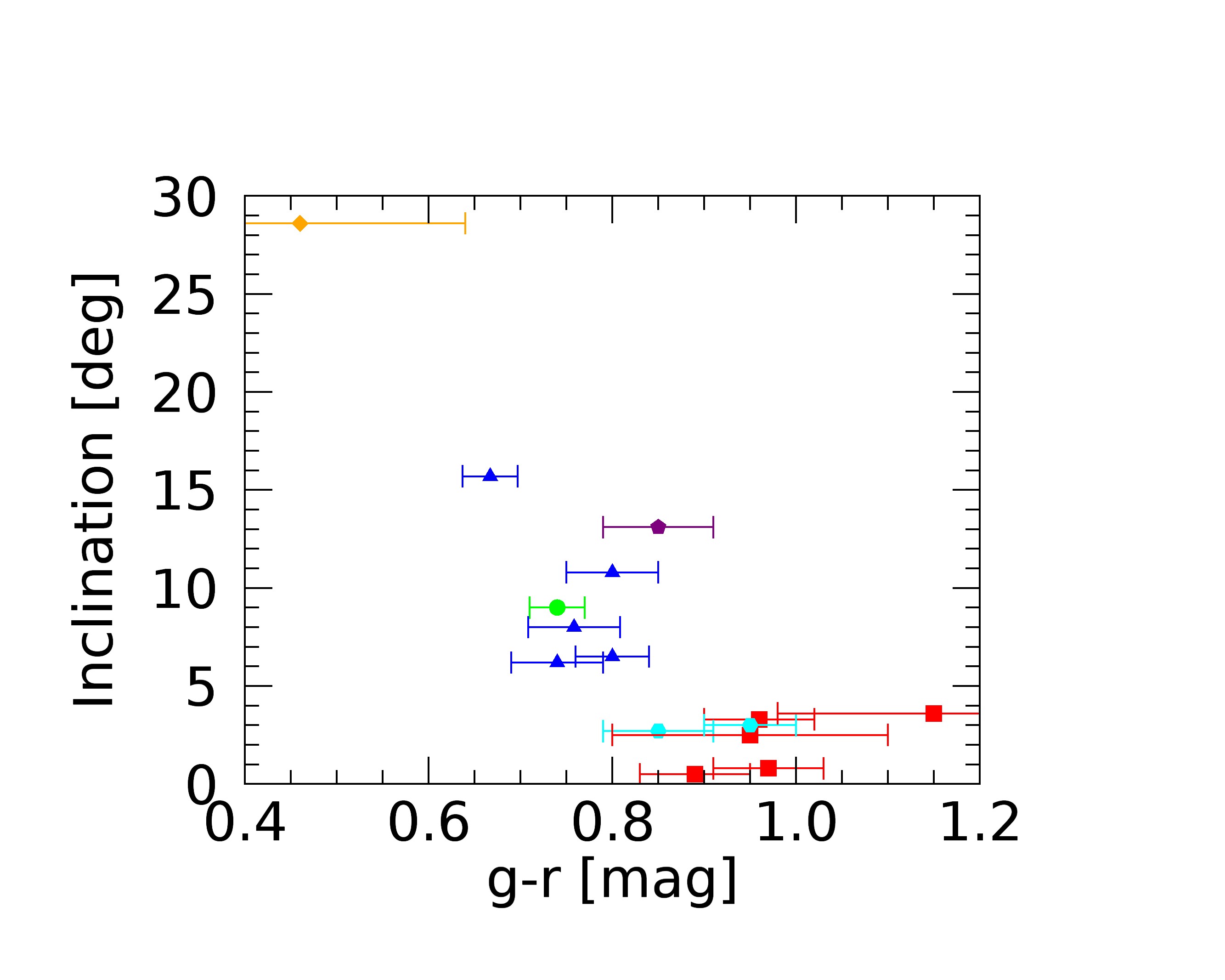}
\includegraphics[width=9cm, angle=0]{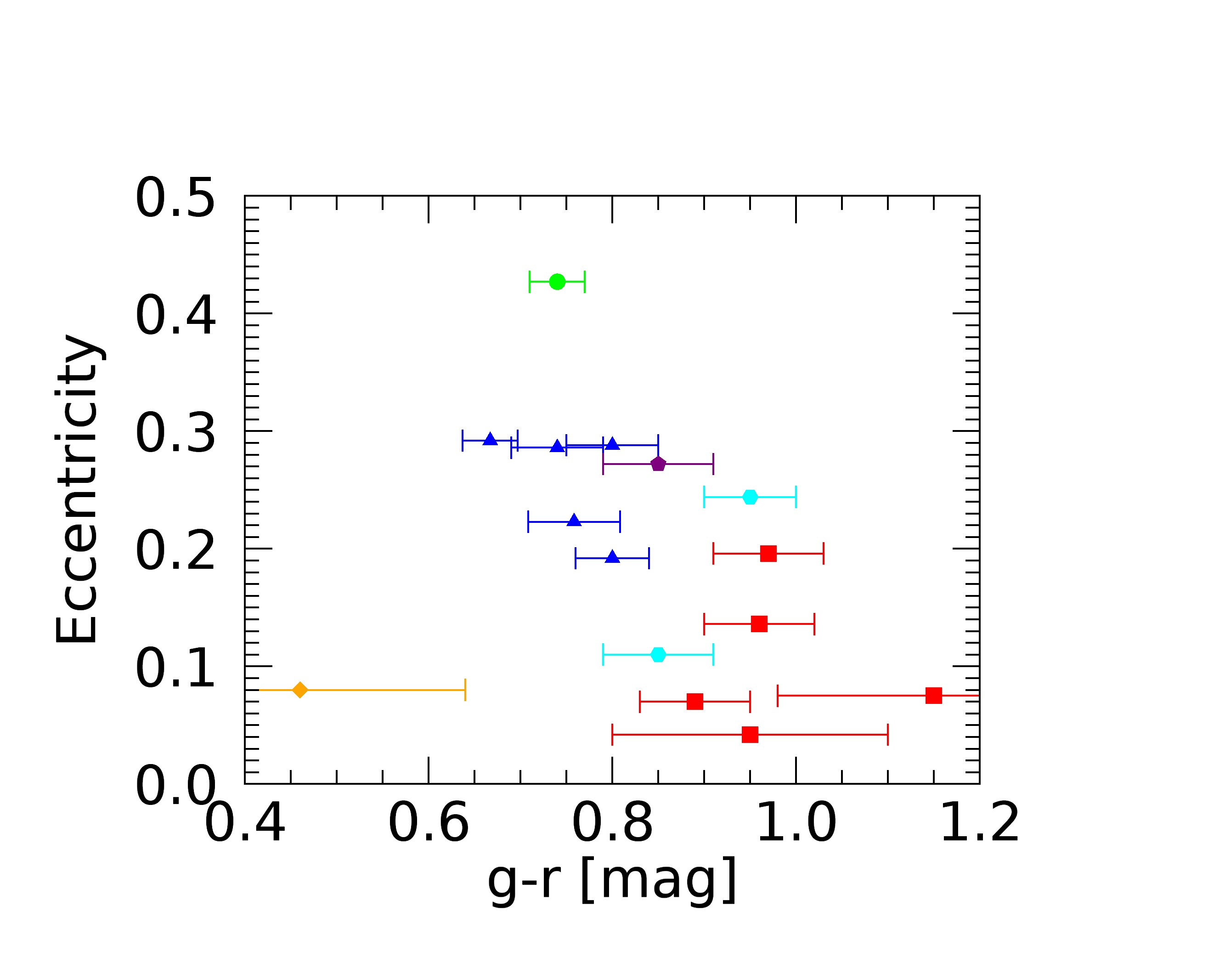}
\includegraphics[width=9cm, angle=0]{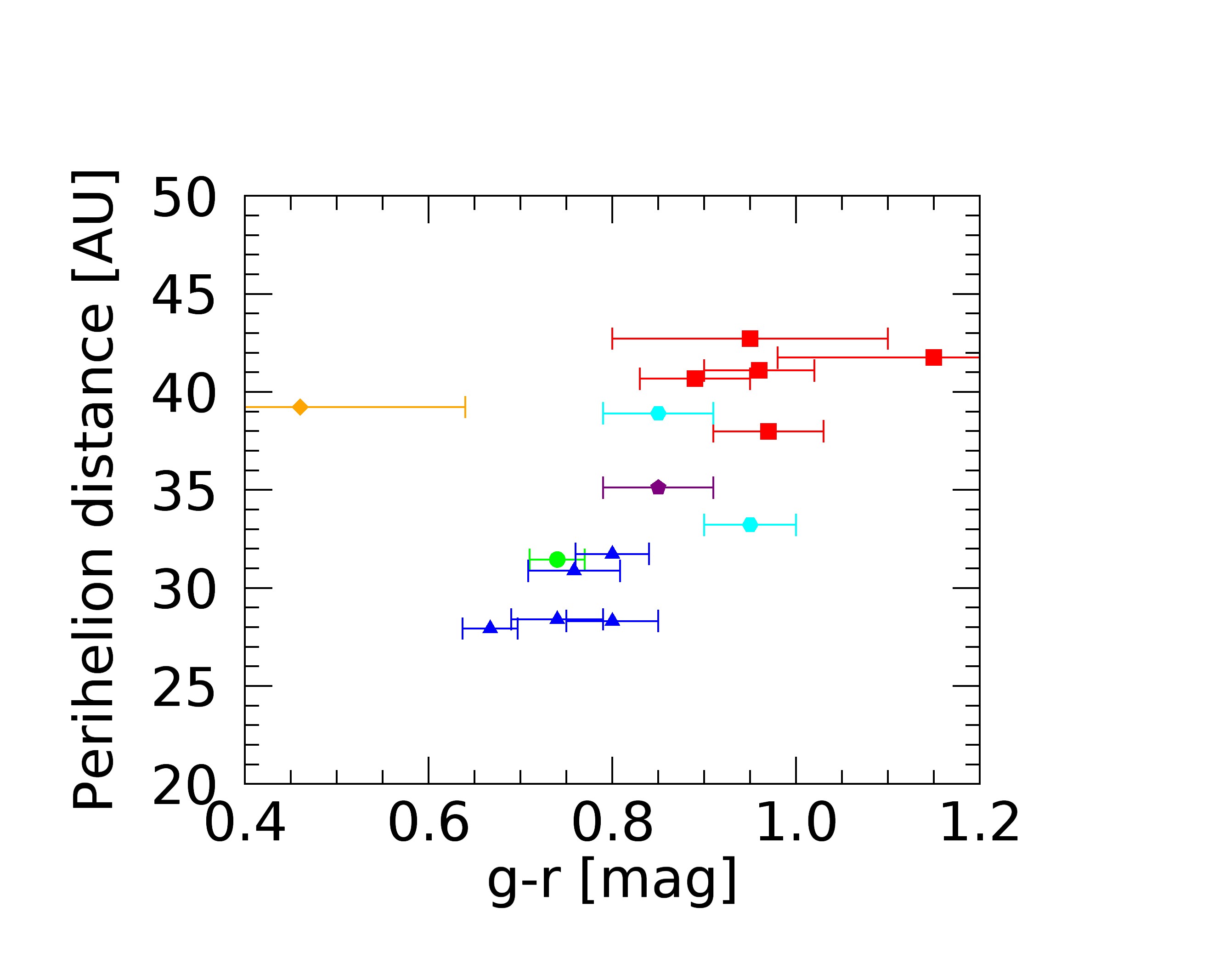}
\includegraphics[width=9cm, angle=0]{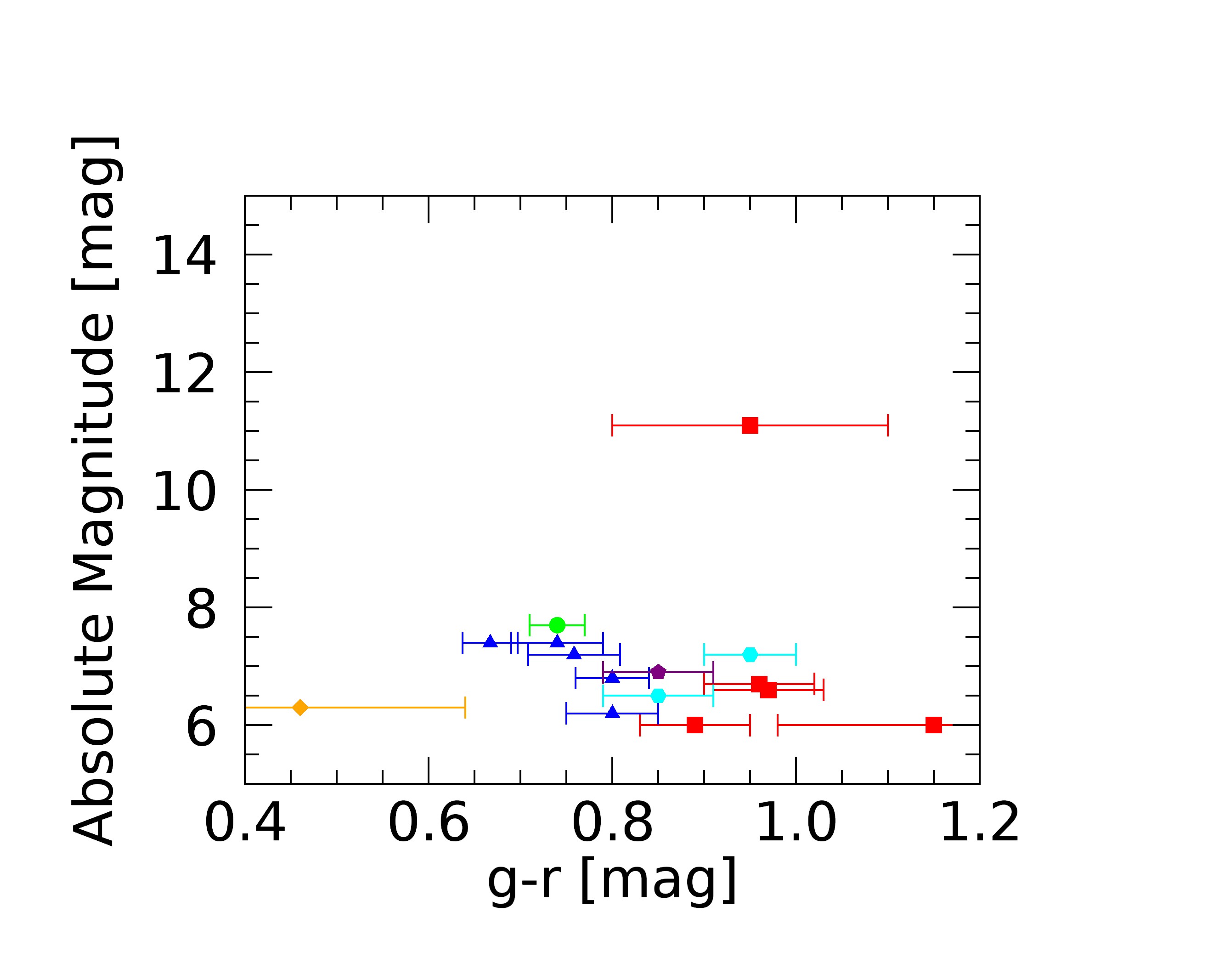}
\includegraphics[width=9cm, angle=0]{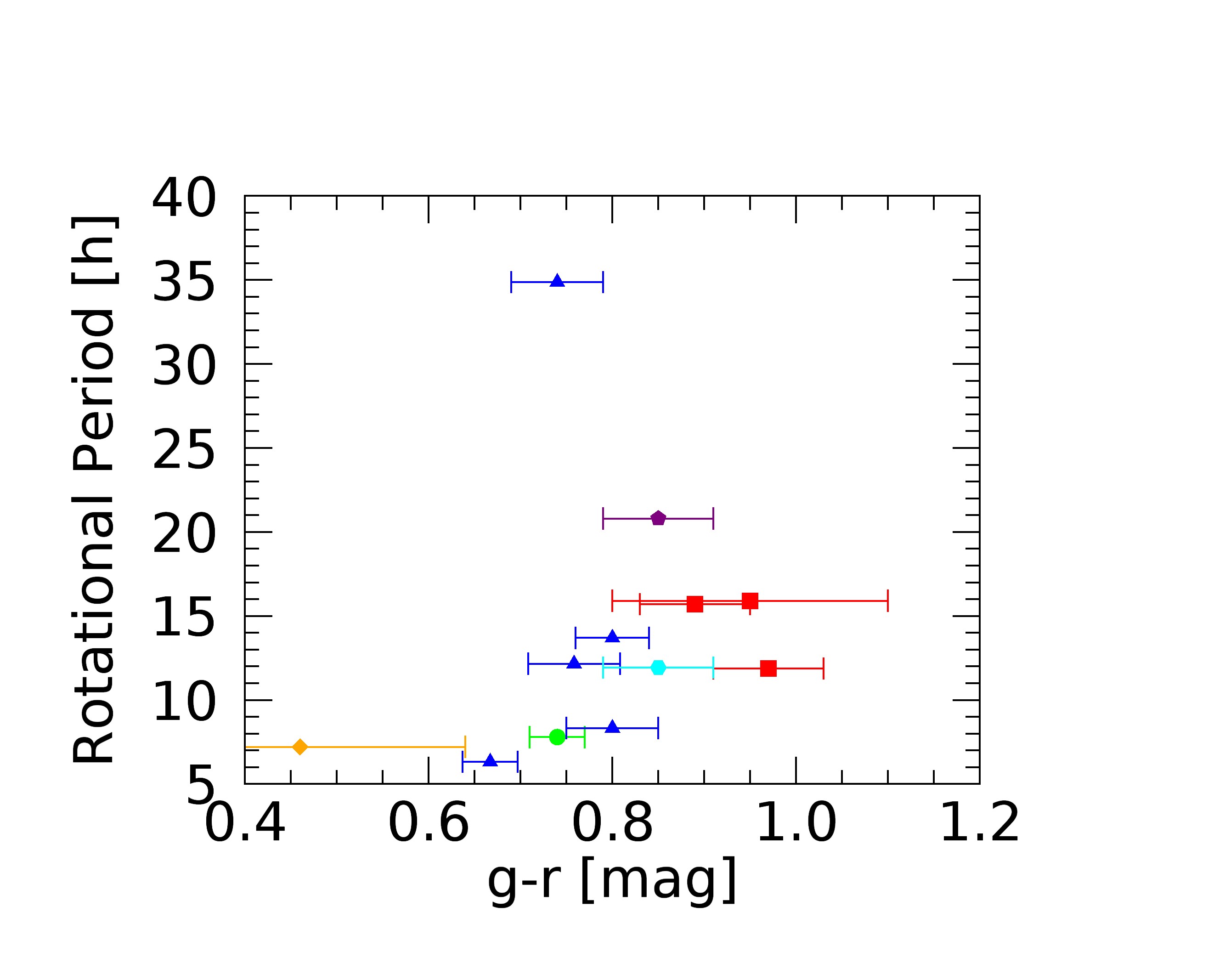}
 \caption{Orbital elements, absolute magnitude and rotational period versus g'-r': Objects discussed in this work are plotted. The legend is: red squares for dynamically Cold Classicals, blue triangles for Plutinos, green circle for 5:2 resonant, cyan hexagon for 7:4 resonants, purple hexagon for 2:1 object and orange diamond for Haumea family member. There is an anti-correlation between inclination and g'-r' and a correlation between rotational period and colors.   }
\label{fig:Incli}
\end{figure*}
 
As mentioned, escaped Cold Classical TNOs have been found in several resonances, generally at low inclination, but some are also at higher inclinations \citep{Sheppard2012}. The very-red/ultra-red resonant contact binaries have inclination from i=2.7 to 13.1$^\circ$ whereas the moderately-red resonants have i=6.2 to 15.7$^\circ$ (Table~\ref{ObsLog}, Figure~\ref{fig:Color2}). In all cases, resonants discussed in this work have low to moderate inclinations (2.7-15.7$^\circ$). In the case of the Cold Classicals, their inclinations are lower than 3.6$^\circ$ whereas the Haumea family member has the highest inclination with i=28.6$^\circ$.
 
There is an anti-correlation between inclination and g'-r' suggesting that the redder objects are at lower inclination (Figure~\ref{fig:Incli}). Using \citet{Spearman1904} technique, this anti-correlation has a $\rho$=-0.740 and a significance level of 99$\%$ using our entire sample. Without the Haumea family member, we have $\rho$=-0.680 and a significance level of 99$\%$. 
 
 The anti-correlation between inclination and color was first noticed by \citet{Tegler2000} for the Classical population. In fact, the ultra-red material is at low inclination with a cut-off of 5-10$^\circ$ and corresponds to the dynamically Cold Classical TNOs. Several studies looked for similar anti-correlation across the sub-populations as well as the entire belt \citep{Sheppard2012, Peixinho2015, Tegler2016, Marsset2019}. Using the entire belt, the tendency was not obvious but a recent study by \citet{Marsset2019} suggests that such a trend exists. Therefore, the contact binary sample seems to follow the same tendency as the rest of the trans-Neptunian belt.    
  
There is an anti-correlation between eccentricity and g'-r' suggesting that the redder objects are at low eccentricity, but the significance level is low. Using the entire sample, we found $\rho$=-0.505 with a significance level of 94$\%$, Without the Haumea family member (2003~SQ$_{317}$), we estimated $\rho$=-0.706 at 99$\%$. Also, there is a potential correlation between colors and perihelion distance. With the entire sample, the correlation has a $\rho$=0.593 for a significance level of 97$\%$, but without 2003~SQ$_{317}$ we found $\rho$=0.815 at 99$\%$. There is no evidence for strong correlation/anti-correlation between colors and perihelion distance and eccentricity among the entire trans-Neptunian belt \citep{Peixinho2015}. Therefore, it is unclear if these trends are characteristic of the contact binary population or if we are dealing with an observational bias due to our limited sample.

Rotational periods have been derived for most of the confirmed and likely contact binaries. Only 2004~VU$_{75}$, 2004~MU$_{8}$ and 2013~FR$_{28}$ do not have a rotational lightcurve yet. We found a potential correlation between colors and periods suggesting that the redder objects are slow rotators. Using our entire sample, we computed $\rho$=0.471 at 88$\%$ but without 2014~JL$_{80}$ (the slowest rotator), $\rho$=0.680 at 97$\%$ (Figure~\ref{fig:Incli}). Colors are an indicator of age: bluer objects have younger surfaces whereas the redder ones are primordial older surfaces. Collisions can expose icy material and thus resurface an object making it bluer as well as affect its rotation. Assuming that collisions spin up the rotation, the bluer objects are the youngest, are fast rotators and have suffered a strong collisional history. However, \citet{Thirouin2016} inferred that the Haumea family members (i.e. blue objects) tend to rotate faster than the other TNOs and \citet{Thirouin2019} suggested that the dynamically Cold Classical TNOs (i.e., very-red/ultra-red objects) tend to rotate slowly. In both studies, we argued that these differences are likely due to the formation/evolution of these objects. It also seems that the escaped Cold Classicals identified as likely contact binaries are also rotating slower than the rest of the TNOs, as the rest of the dynamically Cold Classical TNOs \citep{Thirouin2019}. 

Finally, we also looked for any trend between colors and absolute magnitude (Figure~\ref{fig:Incli}). But, there is no obvious trend. Similarly, there is no trend between colors and the other orbital elements. 

 \subsection{Implication for the formation of contact binaries}
 
The contact binary population does not display any significant differences in color compared to the rest of the TNOs (based on our limited sample). 
 
In \citet{Thirouin2018, Thirouin2019}, we derived that 40-50$\%$ of the Plutinos and only 10-25$\%$ in the dynamically Cold Classical population could be nearly equal-sized contact binaries. There is no estimate for the 2:1, 7:4 and 5:2 resonances yet, but if they are similar to the 3:2, we may expect a large number of contact binaries. Also, it seems that the resonances are a good place to look for contact binaries as more than half of them have been found in resonances.
 
The formation of contact binaries is still an open question. Four models have been proposed: i) Kozai and other dynamical effects can shrink the orbit of wider binaries \citep{Porter2012}, ii) fragmentation during gravitational collapse of a cloud of particles during formation \citep{Nesvorny2010}, iii) three-body type interactions \citep{Goldreich2002}, and iv) gentle collision between two objects \citep{Stern2019}. Because the dynamically Cold Classical population never suffered any strong collisional evolution, the gentle collision process seems to be an adequate option, but we cannot discard the other options. However, because resonances have a different formation and evolution, the gentle collision seems unlikely to explain the creation of contact binaries in resonances. These differences in dynamical history also likely account for the difference in the wide binary fraction seen for the Cold Classicals compared to the resonances and scattered disk objects (see \citet{Thirouin2019}). That is, the more significant dynamical interactions likely experienced by the resonance and scattered objects caused most wide binaries to be unstable, with the wide secondary either being lost from the system or possibly collapsing down to form a closer or contact binary (see also \citet{Thirouin2019} and \citet{Nesvorny2019}).
 
Despite finding a handful of contact binaries in the dynamically Cold Classicals, we have hints, based on their ultra-red/very-red colors and low/moderate inclination, that several escaped Cold Classicals are contact binaries. In fact, there are ten ultra-red/very-red (five in resonances and 5 in the Cold Classical population) and five moderately-red/neutral contact binaries. Therefore, despite the limited sample, it seems that some of the contact binaries in resonances are linked (or potentially) to the Cold Classicals. 

For the escaped Cold Classical contact binaries, it is necessary to discuss where they have been formed. The contact binary could have been formed in the dynamically Cold Classical population and escaped as contact binary or it could have been formed once in resonance (or during its escape). As the dynamically Cold Classical population is dominated by resolved wide binaries it is not unrealistic to expect that resolved wide binaries were able to escape their main reservoir and are now trapped in resonances. With the discovery of a wide binary with ultra-red color in the 3:2 population ((341520) Mors-Somnus, 2007~TY$_{430}$), the feasibility of this scenario has been demonstrated \citep{Sheppard2012b}. Then, through dynamical processes and/or tidal effects, these wide binaries can shrink their orbits and end up as contact binaries. On the other hand, if a wide-binary like 2007~TY$_{430}$ is able to conserve its binarity after escaping the dynamically Cold Classical population despite the fact that wide binaries are very sensitive to perturbation, a contact binary in a such compact configuration will keep its binarity \citep{Parker2011}.

Interestingly, based on our survey of Cold Classical lightcurves, we are not finding a lot of contact binaries in this population as in the Plutino one \citep{Thirouin2018, Thirouin2019}. Assuming that escaped Cold Classical classified as contact binaries were formed in their main reservoir, we should find a lot of contact binaries in the Cold Classical population. Therefore, it seems more likely that the escaped Cold Classical contact binaries were formed after or during their escape from the main reservoir. The different dynamical evolution of the resonant TNOs compared to the Cold Classical population may be a clue to understand the contact binary formation.


\section{Summary and Conclusions}  

We report new color measurements for seven likely and potential contact binaries in the Kuiper belt. Our results are: 
\begin{itemize}
\item All the potential and likely contact binaries in the dynamically Cold Classicals display very-red/ultra-red surface colors. These colors are typical for this sub-population and thus they have likely formed in-situ. 
\item The resonant trans-Neptunian objects reported in this work display a variety of colors from moderately-red to ultra-red. Four have very-red color and one has ultra-red color, making them potentially escaped Cold Classicals. Four have moderately red surfaces and thus are not linked to the dynamically Cold Classical population. As moderately-red surface colors are common in several sub-populations of TNOs, we cannot pinpoint their origin, but they were likely scattered to their current location.
\item There is a strong anti-correlation between inclination and g'-r'. Similar trend is noticed in the entire Kuiper belt. We also present a potential correlation between perihelion distance and color, between rotational period and colors as well as an anti-correlation between eccentricity and color. It is unclear if these trends are real or are due our too limited sample. Finding more contact binaries will help to confirm these trends.
\item For the escaped Cold Classicals classified as likely contact binaries, they can form in the main reservoir of Cold Classical or once trapped in resonances. It is possible that resolved wide binaries escaped their main reservoir and through dynamical processes they shrunk their orbit and are now an end-state binary system in a contact configuration. A second option is that contact binaries were formed in the dynamically Cold Classical population and escaped as contact binaries. Because we are not finding a lot of contact binaries in the Cold Classical population, and because so far the resonances have been the most prolific, we suggest that the formation of contact binaries happened from dynamical interactions during or after their escape from the main reservoir. 
\end{itemize}

\acknowledgments

Authors thank the referee for her/his careful reading of this paper as well as comments. This paper includes data gathered with the 6.5~m Magellan-Baade Telescope located at las Campanas Observatory, Chile. Authors acknowledge the Magellan staff. Authors also acknowledge support from the National Science Foundation (NSF), grant number AST-1734484 awarded to the ``Comprehensive Study of the Most Pristine Objects Known in the Outer Solar System".

\clearpage 

\end{document}